\documentclass[numberedappendix,twocolumn,apj]{emulateapj}

\shorttitle{Evolution of Merged Stars}
\shortauthors{Suzuki et al.}

\begin{document}
\title{Evolution of Collisionally Merged Massive Stars}

\author{Takeru K. Suzuki$^{1}$, Naohito Nakasato$^{2}$, 
Holger Baumgardt$^{3}$, Akihiko Ibukiyama$^{2}$, Junichiro Makino$^{4}$
\& Toshi Ebisuzaki$^{2}$}
\email{stakeru@ea.c.u-tokyo.ac.jp}
\altaffiltext{1}{Graduate School of Arts and Sciences, University of Tokyo,
Komaba, Meguro, Tokyo, Japan, 153-8902}
\altaffiltext{2}{RIKEN, 2-1 Hirosawa, Wako 351-0198, Japan}
\altaffiltext{3}{AIfA, Universit\"{a}t Bonn, Auf dem H\"{u}gel 71, 
53121 Bonn, Germany}
\altaffiltext{4}{Division of Theoretical Astrophysics, National Astronomical 
Observatory of Japan, Mitaka, Tokyo, Japan, 181-8588}

\begin{abstract}
We investigate the evolution of collisionally merged stars with mass of 
$\sim 100 M_{\odot}$ which might be formed in dense star clusters.  
We assumed that massive stars with several tens $M_{\odot}$ collide 
typically after $\sim 1$Myr of the formation of the cluster and 
performed hydrodynamical simulations of several collision events. 
Our simulations show that after the collisions, merged stars have extended envelopes
and their radii are larger than those in the thermal equilibrium states and 
that their interiors 
are He-rich because of the stellar evolution of the progenitor stars.
We also found that if the mass-ratio of merging stars is far from 
unity, the interior of the merger product is not well mixed and the 
elemental abundance is not homogeneous.   
We then followed the evolution of these collision products by a one 
dimensional stellar evolution code.  
After an initial contraction on the Kelvin-Helmholtz (thermal adjustment) 
timescale ($\sim 10^{3-4}$ yr), the evolution of the merged stars traces 
that of single homogeneous stars with corresponding masses and abundances, 
while the initial contraction phase shows variations which depend on the mass 
ratio of the merged stars.    
We infer that, once runaway collisions have set in, subsequent collisions of 
the merged stars take place before mass loss by stellar winds becomes significant.
Hence, stellar mass loss does not inhibit the formation of 
massive stars with mass of $\sim 1000M_{\odot}$. 

\end{abstract}

\section{Introduction}

Recent infrared observations of the Galactic center and the centers of other nearby
galaxies have revealed a population of compact and massive star clusters 
located 
close to the centers of galaxies, such as the Arches and Quintuplet 
clusters \citep{oku90, fig99, fig02}, 
IRS 13E \citep{mai04}, and IRS 16SW \citep{lu05} and MGG-11 in M82 \citep{mkk03}. 
The estimated masses of these clusters are in the range of $10^4 M_\odot$ 
to $10^5 M_\odot$ 
while their half-mass radii are between 0.1 and 1 pc, giving rise to central densities in excess 
of $10^6 M_\odot/pc^3$. 

Dynamical simulations have also shown that if star clusters are born with 
sufficiently high central density, massive 
stars with $>20 M_\odot$ will sink to the cluster center within a few Myr,  
i.e., before the end of the stable nuclear burning phase, through 
dynamical friction \citep{pzm02, pzetal04, frei06a, frei06b}.
Their stellar radii are large enough that there is a high chance for collisions
between them after the stars arrived in the center.


Indeed, $N$-body and Monte Carlo simulations have shown 
that collisions between high-mass stars in young star clusters can lead 
to the formation of a supermassive star with mass of several hundreds
to several thousands $M_\odot$ \citep{pzetal04, bau06, frei06b}.
Such supermassive stars and the intermediate-mass
black holes (IMBHs) which might form out of them could be 
the ultra-luminous X-ray sources recently discovered by {\it Chandra} 
and HST observations \citep{hop04, bau06, pat06}.
It is also argued that runaway massive stars that are ejected from dense 
clusters become a candidate of gamma-ray bursts \citep{ham06}.  

However, whether supermassive stars can really form through runaway collisions
and whether IMBHs forms at the end of their lifetime is still not clear. 
So far, most
simulations have neglected hydrodynamical processes during the collisions and 
the effects of stellar evolution. 
Stars formed from the merging of other stars might start their lives 
with significant abundance gradients because of incomplete mixing. 
In addition, since merging events happen only after a star
cluster has gone into core-collapse \citep{pzetal04, frei06b}, runaway stars 
initially have higher He abundance as a result of the nuclear burning of 
the parent stars. Evolution of stars formed through merging, thus, is 
likely to be different from the evolution of homogeneous stars with ``normal'' 
abundance. 


It has been argued that the formation of IMBHs
from metal-rich ($\sim$ solar abundance)
massive stars is unlikely,
because strong stellar winds \citep{kud02, nug00} considerably reduce the 
masses before the black holes form \citep{bel07,yun06}.
However, 
because the lifetime of merged stars with higher He content is shorter than 
that of normal stars, the total mass lost during the lifetime might be  
smaller than these estimates.


The present paper is a first attempt towards a realistic treatment of the stellar 
evolution of runaway 
stars. In the present paper we will follow the collision of two stars by means of SPH calculations and then 
follow the evolution of the merger product with a stellar evolution code.
The parameters of the colliding
stars are taken from the results of $N$-body simulations of
runaway merging of stars in young star clusters \citep{pzetal04, bau06}.

\section{Method}
\label{sec:met}
Our procedure consists of three steps : (1) stellar evolution of single stars,
(2) simulations of stellar collisions and (3) stellar evolution of collision products. 
This procedure is essentially same as that used by \cite{Sills_1997,Sills_2001}.
In their works, they concentrated on the formation of blue straggler stars
due to the collision between low mass stars ($< 1 M_{\odot}$). 
However, in the present paper we are interested in the
merging process and subsequent stellar evolution in the core of a very dense 
star cluster
where only massive stars ($> 10 M_{\odot}$) are involved in collisions because 
of mass segregation.

\citet{pzetal04} showed that collisions in very dense clusters typically start 
at $t\sim 1$Myr after their formation.    
Following this result, we consider various collisions (Table \ref{tab:sum}) 
of massive stars after $t=1$ Myr from the formation. 
We determine the interior structure of the merged stars by 
calculating the stellar evolution of single stars with solar abundances 
from zero-age main sequence (ZAMS) phase (Step (1) of the procedure). 

The stellar evolution is handled by one-dimensional (1D) spherical symmetrical 
stellar evolution code. Our code, which adopts a usual Henyey method, is 
based on the program originally developed by \citet{pac70}. 
As adopted in general stellar evolution calculation, we neglect 
hydrodynamical evolution and only treat evolution on a 
Kelvin-Helmholtz (thermal adjustment) timescale, 
\begin{eqnarray}
\hspace{-5mm}
\tau_{\rm KH} &\approx& \frac{G M_{\star}^2}{2R_{\star}L_{\star}}  \nonumber \\
&=&10^3\;{\rm yr} \left(\frac{M_{\star}}{100M_{\odot}}\right)^2 
\left(\frac{R_{\star}}{50R_{\odot}}\right)^{-1}
\left(\frac{L_{\star}}{10^{6.5}L_{\odot}}\right)^{-1}, 
\label{eq:KH}
\end{eqnarray} 
where $G$ is the gravitational constant, and $M_{\star}$ ($M_{\odot}$), 
$R_{\star}$ ($R_{\odot}$), and $L_{\star}$ ($L_{\odot}$) are stellar (solar) 
mass, radius, and luminosity, respectively.  
Our code adopts the OPAL opacity \citep{ir96} and equation of state \citep{rsi96} 
tables, and the nuclear reaction rates tabulated in \citet{bu88}.  
We focus on the early phase of stellar evolution before the central H 
is exhausted. 
For this purpose it is reasonable to switch off nuclear burning of He 
and heavier elements, and to consider 
only H-burning because the central temperature is still not high 
($<5\times 10^7$K in our simulations).   

For Step (2), we construct spherical symmetric three dimensional (3D) stellar 
structures from the 1D results.
Then, we perform 3D smoothed-particle-hydrodynamical (SPH) 
simulations of stellar collisions with the parameters summarized in 
Table \ref{tab:sum}.  
We use a modified version of the SPH code by \cite{Nakasato_2003}.
An important modification is that we include a treatment of
equation of state (EOS) that takes into account radiation pressure in 
addition to a fully ionized ideal gas ($\gamma = 5/3$).
We adopt a Balsara type artificial viscosity \citep{Balsara_1995}
with viscosity parameters $\alpha = \beta = 5/6$ as suggested by 
\cite{Lombardi_2003}. 

For the equal mass cases (EQ1 and EQ2 in Table \ref{tab:sum}), 
we use $N = 10,000$ particles for each star (mass resolution of $8.85 \times 10^{-3}$ $M_{\odot}$).
For the unequal mass cases (UE1 and UE2), we use $N = 20,000$ particles for Star 1 
and $N = 6305$ for Star 2, respectively (mass resolution of $4.42 \times 10^{-3}$ $M_{\odot}$).
As will be discussed in 
Appendix this relatively small $N$ is sufficient for our purpose
and our results of SPH simulations do not strongly depend on $N$. 

We assume that initially each star is separated by impact parameter, 
$2\times(R_1 + R_2)$, in all the cases, where $R_1$ and $R_2$ are the stellar 
radii before the collision. We put the first star at the origin and the 
other at the x-axis with a specific tangential velocity as shown in Table 
\ref{tab:sum}. These velocities are typical for collisions that occur 
in central regions of dense clusters \citep{pzetal04}. 
Larger initial velocities lead to less eccentric orbits and larger pericenter distances 
between the two stars and therefore longer merging times.
During SPH simulations, we always check whether merging occurs or not
by assigning each particle into a membership either 1st star,
2nd star or unbound using enthalpy of each particle \citep{Rasio_1991}.
We stop a SPH run at least $\delta t_{\rm merging} = 2 \times 10^5$ sec 
after the mering.
Note that the choice of $\delta t_{\rm merging}$ is arbitrary and does not 
affect the results as long as it is sufficiently longer than the dynamical 
time scale.
This criterion ensures that the particles are in dynamically stable states.  
The last snapshot is used to create the structure of a merged star that is 
evolved in Step (3).

In Step (3), we follow the evolution of the collision products by  
the same 1D stellar evolution code used in Step (1). 
To do this, we need to construct 1D structure from the results of 
3D SPH simulations in Step (2).   
In this paper, we neglect the effect of rotation, and simply average 
radius, $r$, density, $\rho$, temperature, 
$T$, and elemental abundances of the 3D results in mass 
radius coordinate, $m$ to give 1D spherical symmetric structure.  
Needless to say, this treatment is a very simplified one. 
The SPH simulations show that the end-products of the merging events are 
not spherical due to the rotation (see Figure \ref{fig:SPH1}).
For example, the ratio between the final rotational velocity and
the circular velocity at a given radius ranges between
0.2 (inner region) and 0.8 (outer region).
Indeed, for more detailed studies, we should adopt a more elaborated way 
that takes into account rotation when mapping a 3D distribution into a 1D 
spherical symmetric profile \citep[e.g.,][]{Sills_2001}.
However, since the rotational energy of the merged stars is less than 15 \% 
of the gravitational
energy, the rotation is not expected to affect the evolution of the stars much.
Thus, we think that we can give rough but reasonable estimates for the 
evolution of the collision products by our simple prescription. 

On the other hand, the SPH simulations cannot treat low density envelopes 
on account of limited mass resolution. Then, we extrapolate density and 
temperature structure to the outer region to match the inner structure 
obtained by the SPH simulations. 
In order to do this, we adopt the Eddington approximation to derive the 
relation between temperature and optical depth and assume hydrostatic 
equilibrium to set the density structure in the outer envelopes.

Also, in the SPH simulation we do not consider nuclear energy release, because  
the total nuclear energy production integrated with collision duration 
is much smaller than the gravitational energy. 
In the later evolution of merged stars (Step 3), however, nuclear burning is 
as important as the energy release by gravitational contraction. 
Then, we have to carefully determine luminosity, $l$, from  an energy 
equation, 
\begin{equation}
\label{eq:lmns}
\frac{\partial l}{\partial m} = \epsilon_{\rm n} - \left[\left(
\frac{\partial u}{\partial \rho}\right)_T - \frac{\rho}{p^2}\right]
\frac{\partial \rho}{\partial t} -\left(\frac{\partial u}{\partial T}
\right)_{\rho}\frac{\partial T}{\partial t} ,
\end{equation}
where $\epsilon_{\rm n}$ is net energy gain by nuclear burning minus 
neutrino loss and $u$ is internal energy (e.g. section 9 of 
Kippenhahn \& Weigert 1990). Note that the terms involving 
time derivative denote the energy release (absorb) by gravitation 
contraction (expansion). 
As a matter of calculation technique, we have to set an appropriate 
time-step, $\Delta t$, for these terms in order to proceed 
the stellar evolution calculation stably. 
We adopt $\Delta t \approx (1/100) \tau_{\rm KH} (=10 - 100{\rm yr})$ 
(but much larger than free-fall time scale) just after the merger events to 
precisely follow 
Kelvin-Helmholtz contraction. As merged stars settle down to thermally 
stable state, we adopt larger $\Delta t (\lesssim \tau_{\rm KH})$. 

Because collision products are not in thermally equilibrium states 
just after the mergers, we need to prepare an appropriate initial guess for 
the time evolution of physical variables ($\rho$, $T$, $l$, and $r$ as 
functions of $m$) by the Henyey method;   
without this treatment we fail to follow the evolution of collision products, 
even though an appropriate $\Delta t$ is set.
For an initial guess of the correction, we use a mixture of the structure 
of SPH simulation and thermally relaxed structure (e.q. ZAMS) that is 
determined separately.  
Then, we derive the correct time-evolved structure by relaxation.

\begin{table}
\begin{tabular}{c|cccc}
\hline
  & EQ1 & EQ2 & UE1 & UE2 \\
\hline
Star 1($M_{\odot}$) & 88.5 & 88.5 & 88.5 & 88.5\\
Star 2($M_{\odot}$) & 88.5 & 88.5 & 27.9 & 27.9\\
$e$ & 0.444 & 0.100 & 0.669 & 0.125\\
$V_{\rm init} ({\rm km\; s^{-1}})$ & 550 & 700 & 400 & 650 \\
collision time(day) & 2.1 & 44 & 1.4 & 13 \\
Final Mass ($M_{\odot}$) & 165.5 & 156.4 & 106.1 & 98.0 \\ 
$-dM (M_{\odot})$ (collision) & 11.5 & 20.6 & 10.3 & 18.4 \\ 
\hline
\end{tabular}
\caption{Summary of the four runs. The third and fourth lines show
the orbital eccentricity $e$ and initial tangential velocity of the stars.
The last line gives the mass lost during the collisions ($-dM$).}
\label{tab:sum}
\end{table}

\section{Results}
\subsection{SPH simulation of stellar collisions}
\label{sec:sph}

\begin{figure}
\epsscale{1.0}
\plotone{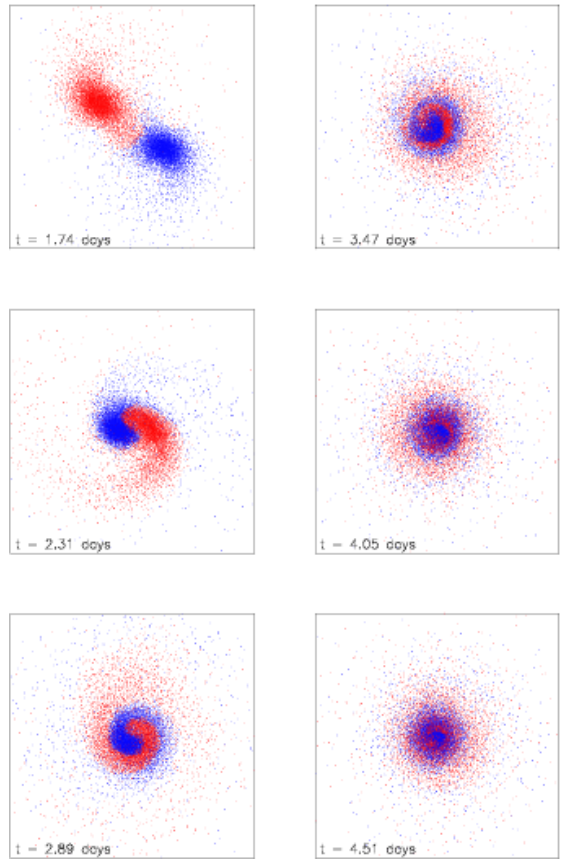}
\caption{
The last phase of the merging process in the SPH simulations for EQ1 case. 
In these snapshots, particles are projected onto the xy plane (orbital plane).
The size of each panel is 100 $R_{\odot}$.
Blue and red points represent the particles originating from Star 1 and 2, respectively.
\label{fig:SPH_SNAP_EQ}}
\end{figure}

\begin{figure}
\epsscale{1.0}
\plotone{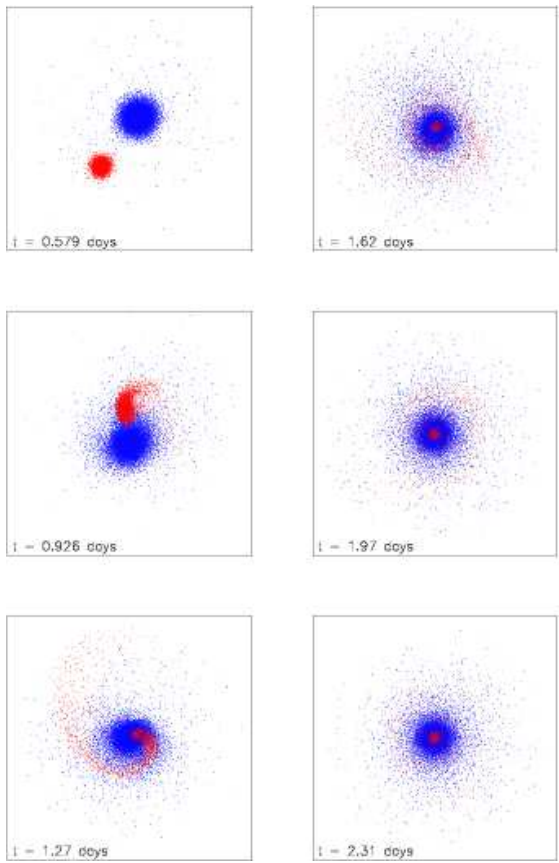}
\caption{
Same as Figure \ref{fig:SPH_SNAP_EQ} but for UE1 case.
\label{fig:SPH_SNAP_UE}}
\end{figure}


In all cases reported in the present work, the overall evolution is qualitatively 
similar: After a certain time, both stars are elongated by tidal interaction
to form an extended merging product as shown in Figures 
\ref{fig:SPH_SNAP_EQ} and \ref{fig:SPH_SNAP_UE}, which depict the last phases 
of the merging processes for cases EQ1 and UE1 respectively.
However, the mass ratio and the orbital eccentricity of the merging stars 
influence the details of the merging processes. 

\subsubsection{Mass Ratio}
\label{sec:msrt}
\begin{figure}
\epsscale{1.0}
\plotone{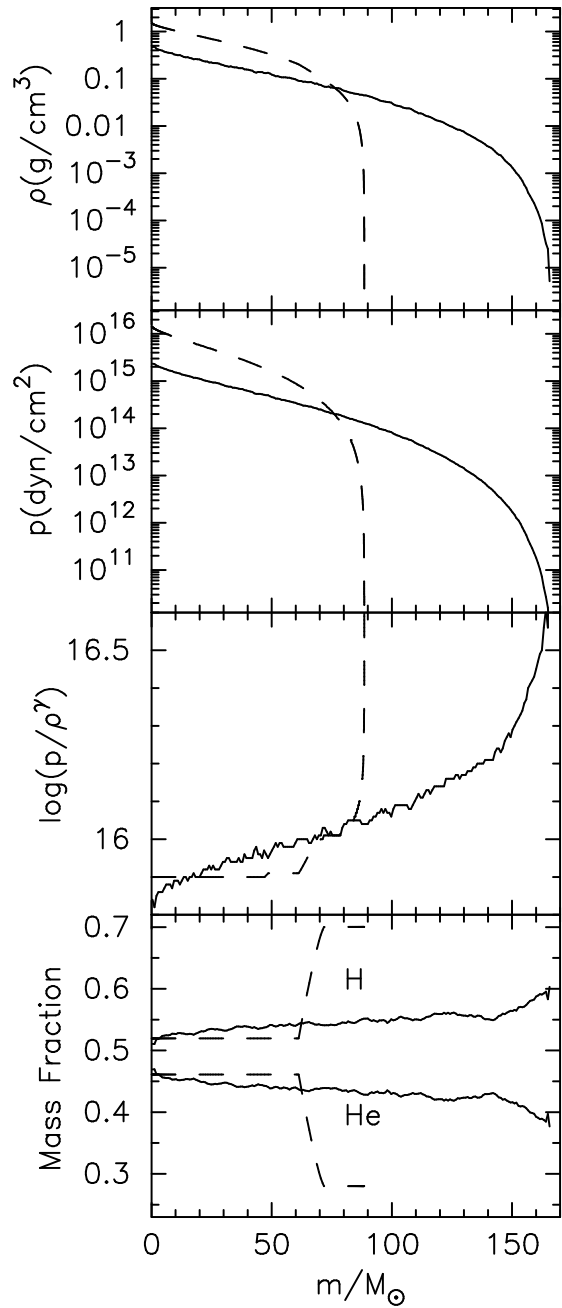}
\caption{Stellar structure of the parent stars (dashed; Star 1$=$Star 2 in 
this case) and the collision product of EQ1 at the end of step 2 (solid). 
The 1D structure of the merged stars are derived from the 3D SPH results 
by averaging in $m$ (see \S\ref{sec:met}). 
From top to bottom, density, pressure, entropy variable, and H and He abundances 
are plotted as functions of mass radius, $m/M_{\odot}$.  
}
\label{fig:str2_eq}
\end{figure}

\begin{figure}
\epsscale{1.0}
\plotone{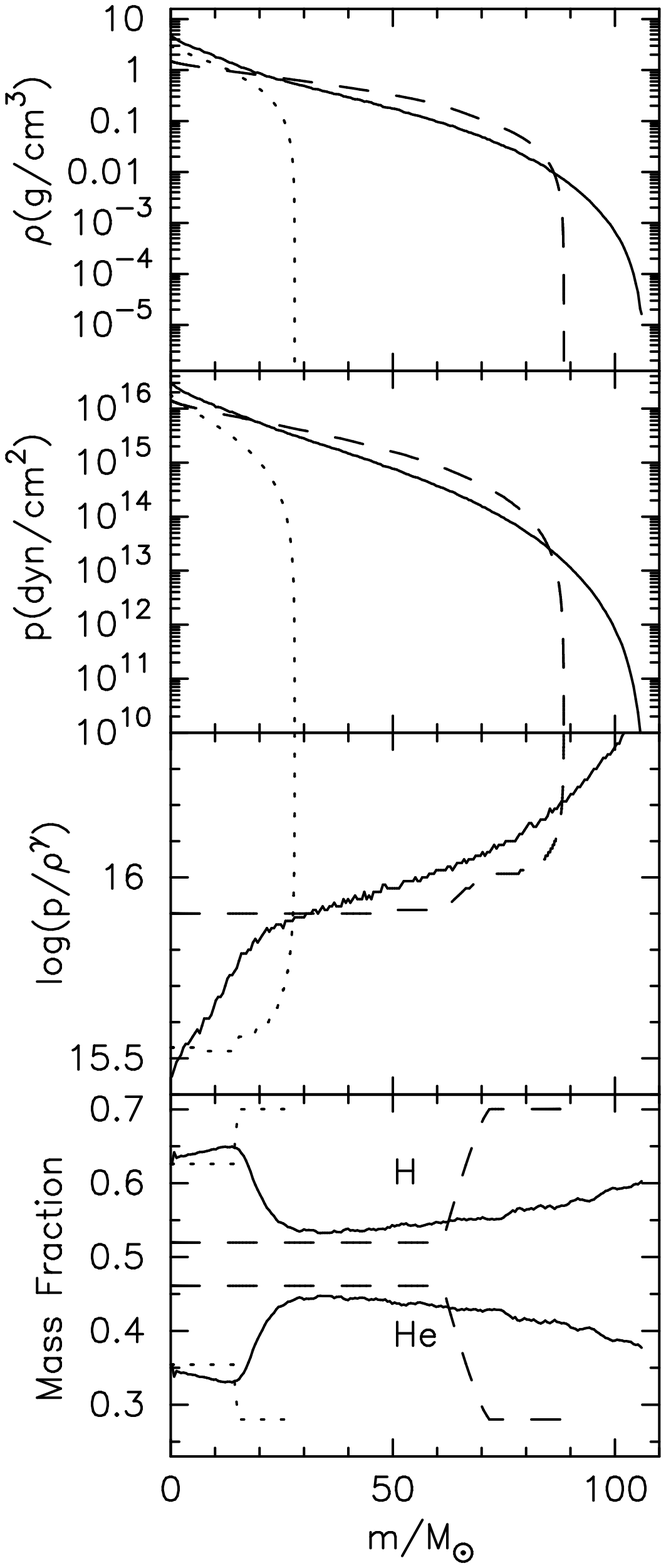}
\caption{The same as Figure \ref{fig:str2_eq} but for UE1. The structure of 
Star 1 and Star 2 are shown in dashed and dotted lines, respectively. 
}
\label{fig:str2_ue}
\end{figure}

The mass ratio affects the material mixing in the interior of 
the merger products. 
Figure \ref{fig:SPH_SNAP_EQ} shows that the material is well mixed in 
the equal mass case, and the elemental abundance is almost homogeneous 
as will be shown later.
On the other hand, in the unequal mass case (Figure 
\ref{fig:SPH_SNAP_UE}) the less massive star (Star 2), which has 
the higher central density, sinks to the 
center without sufficient mixing, and the more massive star (Star 1) 
forms an extended envelope. This is consistent with recent results for 
lower mass stars by \citet{dd06}.  
As a result, the elemental abundance is also 
inhomogeneous in the merger product. This difference of the material mixing 
affects the later evolution of the merged stars, which will be discussed 
in \S \ref{sec:evcp}. 

Let us examine the material mixing during the mergers in more detail. 
Figures \ref{fig:str2_eq} and \ref{fig:str2_ue} respectively compare the 
structure of collision products (solid lines) of EQ1 and UE1 with the 
corresponding parent stars just before the mergers (dashed and dotted lines). 
The 1D structure of the merged stars are reconstructed 
from the 3D results at the ends of the step (2) as explained in 
\S\ref{sec:met}. From top to bottom, density, $\rho$, pressure, $p$, 
an entropy variable, $p/\rho^\gamma$, and H and He abundances are 
plotted against mass radius, $m/M_{\odot}$, where $\gamma$ is a ratio 
of specific heats that is determined from the OPAL equation of state table. 
An entropy variable, $p/\rho^\gamma$, is useful to 
study the material mixing; under the adiabatic condition 
convective stability in uniform 
media with gravity in $-z$ direction reads $d(p/\rho^\gamma)/dz > 0$ 
(Schwartzschild criterion), and in convection zone $d(p/\rho^\gamma)/dz 
= 0$.   Thus, a fluid element with smaller $p/\rho^\gamma$ tends to sink 
into a central region during merger. 

The third panel of Figure \ref{fig:str2_eq} shows that in the 
parent star (Star 1 $=$ Star 2) of EQ1 the large convective core (constant 
$p/\rho^\gamma$) extends to $60 M_{\odot}$. In the He-rich envelope, 
$p/\rho^\gamma$ is larger by only $\approx$ 0.2 dex except the region 
that is very near the surface. Hence, the matter is well mixed during 
the merger because of the low entropy barrier. 
As a result of the mixing,  the gradients of the H and He abundances are very 
small (the bottom panel Figure \ref{fig:str2_eq}).

This result is in contrast to the collision of low mass main sequence (MS) 
stars. 
\citet{lom95} performed the collision of two $0.8M_{\odot}$ MS
stars. They found that the mixing between the dense cores and 
the outer envelopes was inefficient. 
This is because the interior of a low mass star is occupied by a radiative 
core and the fraction of a surface convection zone is tiny in mass. 
Thus, $p/\rho^\gamma$ monotonically increases in the interior; 
e.g. in $0.8M_{\odot}$, $p/\rho^\gamma$ increases by nearly an order of 
magnitude from the center to the envelope \citep{lom02}. 
As a result, the material mixing is inhibited during the merging 
by the large entropy barrier, which gives a clear contrast to our result 
for massive stars.  
 
Next, let us move on to the collision of the unequal mass stars (UE1; Figure 
\ref{fig:str2_ue}). The third panel shows that $p/\rho^\gamma$ of the 
lower mass parent (Star 2; dotted line) is smaller than that of the 
higher mass partner (Star 1; dashed line) because the lower mass star 
has higher central density (the top panel). Owing to the small entropy  
the lower mass star (Star 2) sinks to the core of the merged stars 
without mixing, which was seen in Figure \ref{fig:str2_ue}. 
As a result, the elemental abundances in the core reflect those of Star 2, 
while the abundances in the outer region reflect those of Star 1 (the bottom 
panel); the merged star consists of the H-rich core and the He-rich 
envelope, which is opposed to usual evolved single stars.

\subsubsection{Orbital Eccentricity}

\begin{figure}
\epsscale{1.}
\plotone{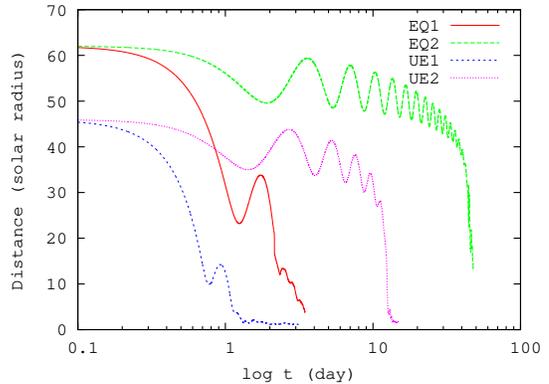}
\caption{
Evolution of the distance between the centers of the merging stars.
In EQ1 and UE1 runs, the two stars merge after a few orbital revolutions,
whereas the EQ2 and UE2 runs need much longer time to merge.
\label{fig:SPH_DISTANCE}
}
\end{figure}

\begin{figure}
\epsscale{1.}
\plottwo{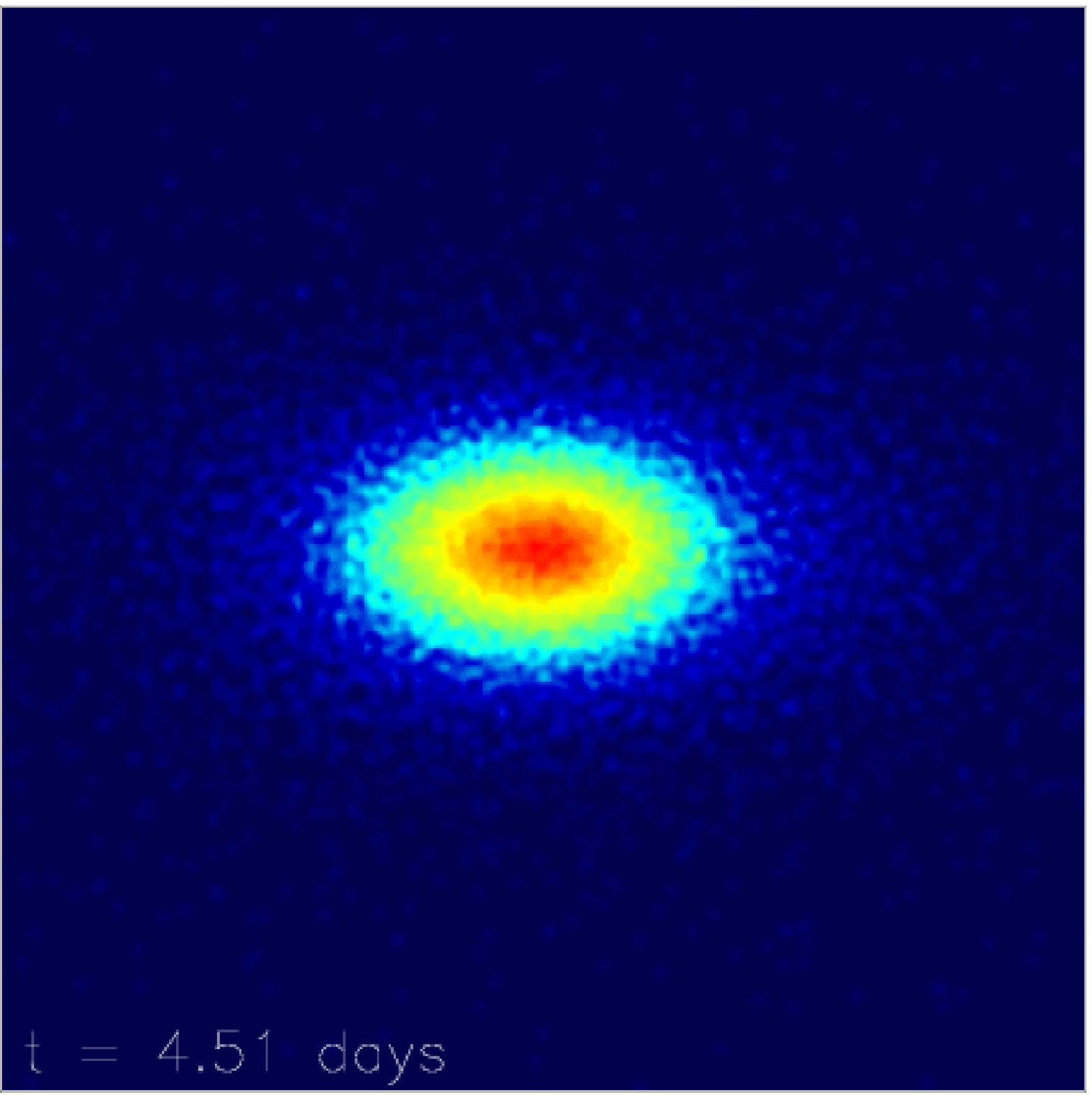}{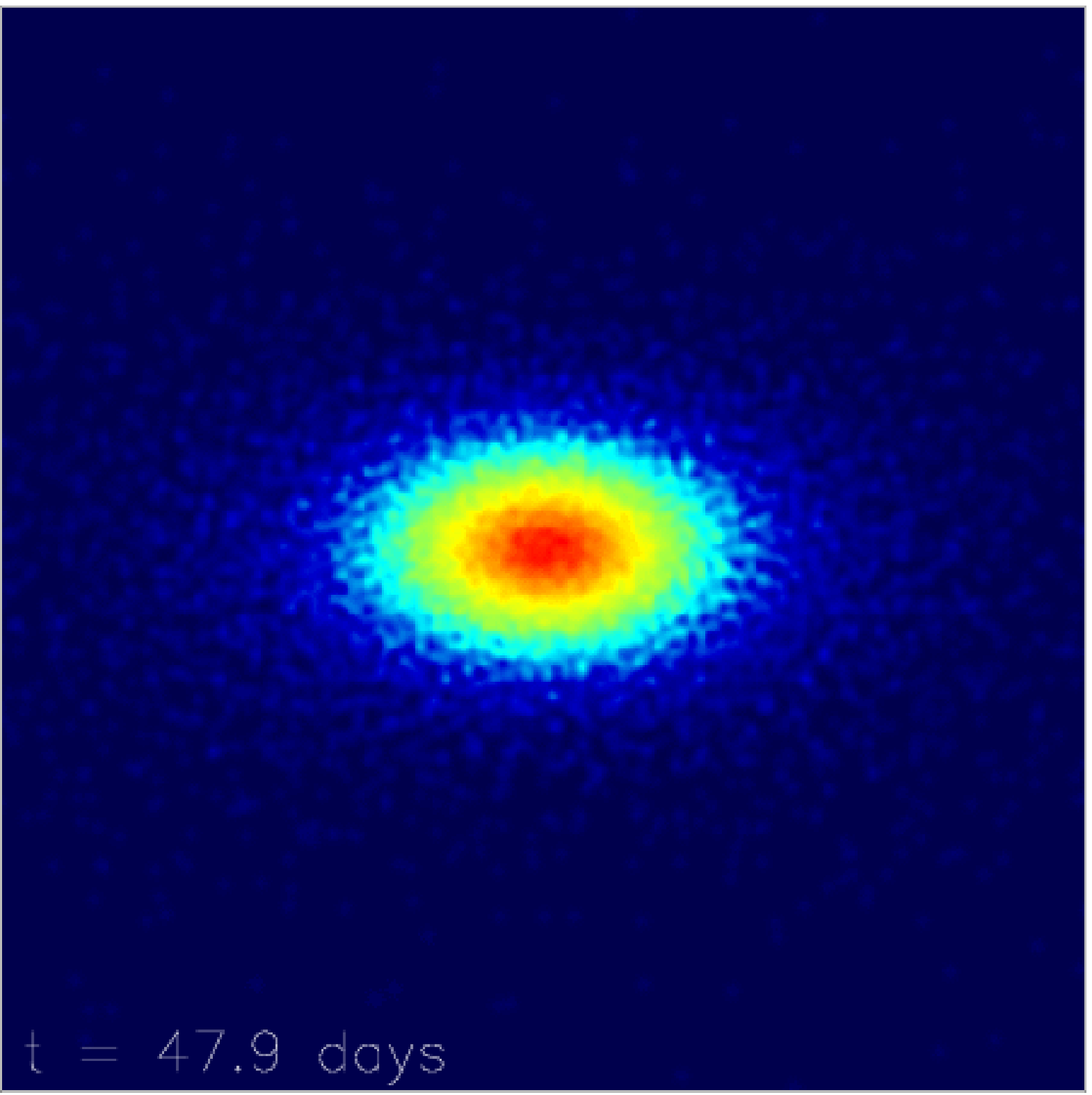}
\caption{
Density snapshots of the merger product at the end of the SPH simulations. 
Particles are projected onto the xz plane where z is the rotation axis.
The size of the panels is 62 $R_{\odot}$. {\it Left} : EQ1 and {\it Right} EQ2}
\label{fig:SPH1}
\end{figure}

The orbital eccentricity controls the time for the stars to merge.  
Figure \ref{fig:SPH_DISTANCE} shows the evolution of the distance between 
the centers of both stars.
In the run with initially more eccentric orbit 
(initially smaller pericenter distance), the stars
merge more quickly than those in more circular orbit, mainly because the 
kinetic energy of the system is smaller; 
EQ1 takes $\sim 2.1 $ days to merge whereas EQ2 takes $\sim 44 $ days.
EQ1 loses only 11.5 $M_{\odot}$ during the merging owing to the shorter 
collision duration, while EQ2 loses a larger mass of 20.6 $M_{\odot}$.   

Despite the different merging time scales, 
the structures of the merged stars, EQ1 and EQ2, are not so different 
after they settle down to dynamically stable states (Figure \ref{fig:SPH1}). 
Therefore, the later evolution of the merged stars is not expected to be 
different between EQ1 and EQ2.  
We found similar tendencies in the unequal mass cases, UE1 \& UE2; in UE1  
(larger $e$), the merger product settles down to a dynamically stable state 
faster and the mass lost during the merger is smaller, while the later 
evolution is essentially the same. 

\subsubsection{Rotation, Difference etc.}
\begin{figure}
\includegraphics[angle=-90,scale=0.3]{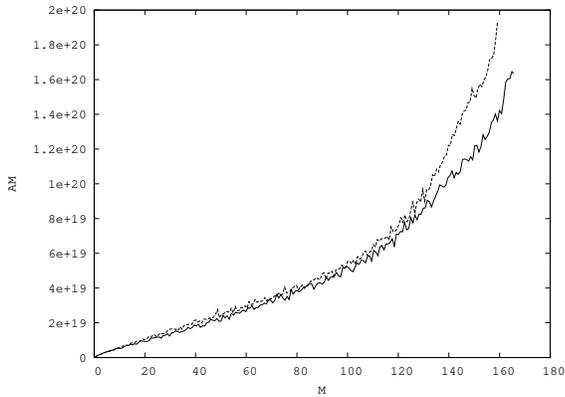}
\caption{
Angular momentum distribution as a function of mass radius for EQ1 (solid line) and EQ2 (dashed line).
}
\label{fig:eqrot}
\end{figure}

As noted previously, the structures of the merged stars, EQ1 and EQ2 (also UE1 and UE2), 
are rather similar but angular momentum (AM) distributions of the merged stars 
are slightly different.
Figure \ref{fig:eqrot} shows AM distribution as a function of mass radius for EQ1 and EQ2.
Clearly, outer AM distribution of EQ2 is larger than that of EQ1 because of the difference in 
the merging time scale.
Namely, during longer orbital revolutions in EQ2, 
more orbital AM is transfered to outer material than in EQ1.
Similar trends are observed in UE1 and UE2 such that
UE2 has slightly larger AM in that outer region than UE1.
Although we neglect the effect of rotation in Step (3) (evolution of 
merging stars)
in the present work, we expect this small difference in outer AM distribution 
gives little influence on subsequent evolution of the merged stars.

\subsection{Evolution of Collision Products}
\label{sec:evcp}
As explained earlier, we follow the evolution of the merged stars by the 
1D stellar evolution code. Since the mass ratio of the merging stars 
affects the structure of the merger products, and since the orbital 
eccentricity does not,  
we mainly study the 
evolution of EQ1 and UE1, and only briefly mention the results of EQ2 and UE2 
for comparison. 
First, we study in detail the stellar evolution 
without taking into account the effect of mass loss by stellar winds. 
Later in \S \ref{sec:ml}, we present the results with mass loss for 
comparison. 

\subsubsection{Equal Mass Collision}

\begin{figure}
\epsscale{1.2}
\plotone{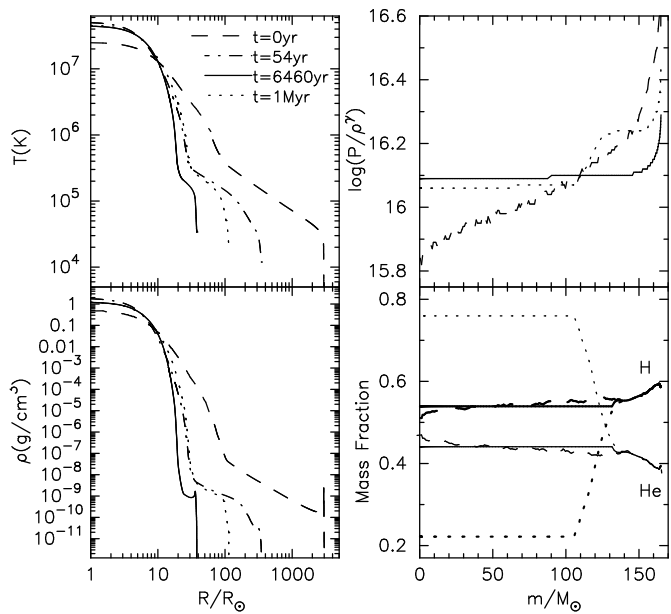}
\caption{Evolution of stellar structure of EQ1. The left panels 
show temperature (top) and density (bottom) as functions of radius in units 
of $R_{\odot}$. The right panels show entropy variable, $p/\rho^\gamma$, 
(top), and H and He abundances (bottom) as functions of mass radius, $m$, 
in units of $M_{\odot}$. 
The dashed, dot-dashed (only for the left panels), dotted, and solid 
lines are the results at $t=0$, 54 yrs, 6460 yrs, and 
1 Myr after the merging, respectively. Note that the radius of the star 
is minimum at $t=6460$ yrs. 
}
\label{fig:str1_rhot}
\end{figure}

\begin{figure}
\epsscale{0.9}
\plotone{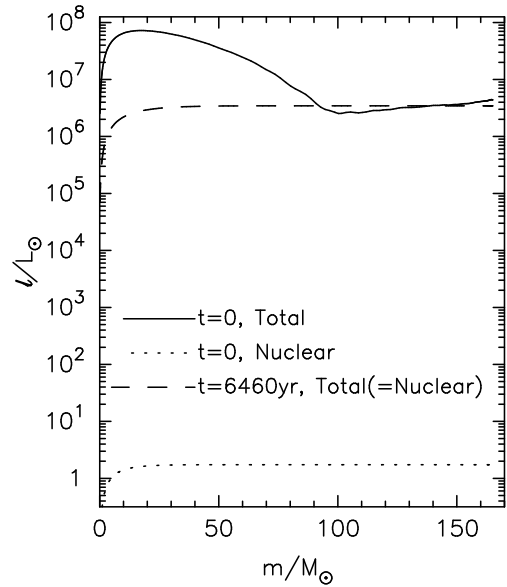}
\caption{
Luminosity normalized by $L_{\odot}$ as a function of $m/M_{\odot}$ for run EQ1. 
The solid and dotted lines give the total luminosity and the luminosity due to the nuclear burning 
at $t=0$. At this stage most of the energy released comes from the gravitational contraction 
of the star. The dashed line shows the total luminosity at $t=6460$ yrs, 
where the energy is coming
from nuclear burning at this time.}
\label{fig:str1_Lr}
\end{figure}


Figure \ref{fig:str1_rhot} shows the evolution of the stellar structure after 
the merging of the equal mass stars (EQ1). 
On the left, temperature (top) and density (bottom) are plotted against 
$r/R_{\odot}$, and on the right entropy variable $p/\rho^\gamma$ (top) 
and elemental abundances are plotted as functions of $m/M_{\odot}$.  
Just after the merger the envelope is extended to $R_{\star}=2960 
R_{\odot}$, exhibiting a core-halo structure owing to the opacity peak 
around $T\simeq 2\times 10^5$K \citep{iuk99} as shown in the left panels 
of Figure \ref{fig:str1_rhot}. 
The bottom right panel shows that the interior is well mixed during the 
merger; the elemental abundance is 
almost homogeneous, $0.52<X<0.6$ and $0.46<Y<0.38$, even at $t=0$, 
where $X,Y$ are H and He abundances.
This is mainly because the entropy gradient is small in the parents stars 
(\S \ref{sec:msrt}).

Figure \ref{fig:str1_Lr} presents luminosity, $l/L_{\odot}$, as a function 
of $m/M_{\odot}$. 
The solid and dashed lines are the total luminosity derived from Equation 
(\ref{eq:lmns}) at $t=0$ and 6460 yrs. 
The dotted line is the luminosity due to only nuclear reaction at $t=0$. 
When we calculate this, we integrate the nuclear 
energy term ($\epsilon_{\rm n}$) on the right hand side of Equation 
(\ref{eq:lmns}). 
Because the central density and temperature are lower than those in the 
thermal equilibrium state at $t=0$, the contribution from 
the nuclear burning to the total luminosity is very small.
Instead, most of the energy comes from the gravitational 
contraction and it is transported outward by convection in the 
large convective core up to mass radius, $m\simeq 110M_{\odot}$;  
the star is in a state similar to that of a pre-main sequence star.     
The total $l$ at $t=0$ (solid line) 
decreases outward between mass radii $20 < m/M_{\odot}< 90$. 
This is because the liberated energy is not converted to radiation but to 
internal energy, namely the increase of the temperature in this region.

\begin{figure}
\epsscale{0.85}
\plotone{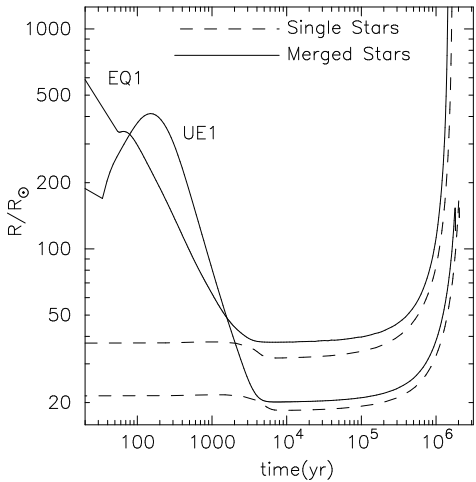}
\caption{Evolution of the radii of the merged stars, EQ1 and UE1 (solid lines), 
in comparison with the chemically homogeneous single star with  
mass, 165.6$M_{\odot}$, and abundance, $(X,Y)=(0.6,0.38)$, and the star 
with 106.1$M_{\odot}$ and $(X,Y)=(0.61,0.37)$ (dashed lines). 
}
\label{fig:t-r1}
\end{figure}

\begin{figure}
\epsscale{1.15}
\plotone{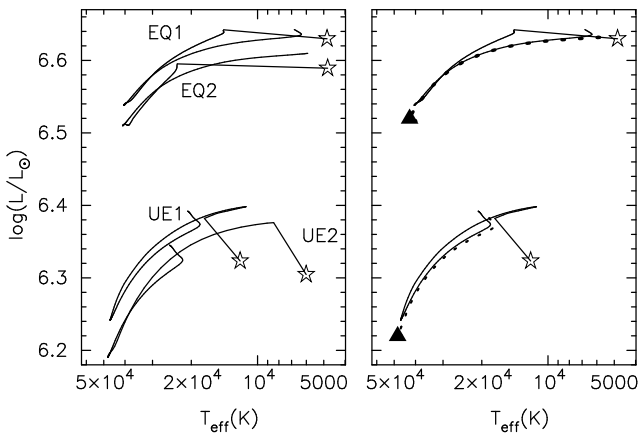}
\caption{Evolution of the merged stars in a HR diagram. In the left panel, 
the solid lines show the results of the four cases. 
Stars are the locations just after the merging.  In the right panel 
the evolution of the chemically homogeneous single stars that are the same 
as in Figure \ref{fig:t-r1} (dotted lines)
in comparison with the 
results of EQ1 and UE1 (solid lines). The triangles are the initial locations 
(ZAMS) of the homogeneous stars. 
}
\label{fig:hr1}
\end{figure}

After the merging, the star contracts towards the thermal 
equilibrium state on a Kelvin-Helmholtz timescale 
(eq. \ref{eq:KH}). 
The temperature and density increase (the left panels of Figure 
\ref{fig:str1_rhot}), and the convective core, which is 
the region with constant $p/\rho^\gamma$, also grows 
until it occupies $\approx 80$\% of the total mass (the top-right panel of 
Figure \ref{fig:str1_rhot}). 
As a result, the elemental abundances become homogeneous inside
$m\lesssim 130M_{\odot}$ (the bottom-right panel of 
Figure \ref{fig:str1_rhot}).
At $t=6460$ yr after the merging, the star contracts to a minimum 
radius, $R_{\star} =37.6R_{\odot}$, after which it expands 
gradually.   
At this time the star is in the stable H-burning phase, and 
the luminosity is all from the nuclear reaction (Figure \ref{fig:str1_Lr}).   
The later evolution traces the evolution of the chemically homogeneous single 
star with the corresponding mass and initial abundance, which we discuss 
below. 

Figure \ref{fig:t-r1} summarizes the evolution of the radii of the merged 
stars (EQ1 as well as UE1; UE1 will be discussed in \S\ref{sec:umcl}), 
in comparison with those of the single stars with the same masses and 
similar abundances ($(X,Y) = (0.6,0.38)$ for EQ1 and $(0.61,0.37)$ for UE1). 
The figure clearly illustrates the monotonical contraction in $t<6460$ yrs, 
which is followed by 
the gradual expansion during MS phase of stable H burning. 
The evolution during the MS phase is similar to that of the single star.  
The increase of the radii at $t\gtrsim 1$Myr indicates the end of MS 
phase due to the exhaustion of H. 
Our calculation shows that the MS lifetime of EQ1 is $\approx 1.3$ Myr, 
which is shorter than the corresponding lifetime ($\approx 2$Myr) of a solar 
abundance star with the same mass due to the smaller initial H abundance.   

Figure \ref{fig:hr1} shows the evolution of the merged stars in a HR 
(Hertzsprung-Russel) diagram; the left panel shows the evolution of the 
four merged stars and the right panel compares the evolution of 
EQ1 and UE1 with the corresponding single stars (for the unequal mass cases, see 
\S\ref{sec:umcl}). 
Through the initial contraction, the effective temperature, $T_{\rm eff}$, of 
EQ1 increases and $L_{\star}$ decreases. The position of the end of the 
Kelvin-Helmholtz contraction (the turning point) in the HR diagram is near 
the ZAMS of the corresponding single stars (triangles). 
The later evolution resembles that of the single stars as we stated previously.
   
The evolution of EQ2 (smaller $e$ than EQ1) is essentially similar to that of 
EQ1 (the left panel of Figure \ref{fig:hr1}). 
Because the mass is slightly smaller, the luminosity becomes lower 
(Figure \ref{fig:hr1}). 
Note that $T_{\rm eff}$ of EQ2 just after 
the stable nuclear burning sets in (the turning point in the HR diagram) is
slightly higher than that of EQ1, although the mass of EQ2 is smaller. This is 
because more massive stars (EQ1) have more extreme core-halo structure 
to show lower $T_{\rm eff}$ (Ishii et al. 1999). 
Chemically homogeneous ZAMS  
stars of which masses exceed a certain limit 
also have this inverse trend. According to Ishii et al. (1999), $T_{\rm eff}$ 
of solar metallicity stars decreases on increasing mass in the range of 
stellar mass $\gtrsim 100M_{\odot}$, while less massive stars show 
the usual trend of the positive correlation between $T_{\rm eff}$ 
and stellar mass. 

\subsubsection{Unequal Mass Collision}
\label{sec:umcl}
\begin{figure}
\epsscale{1.2}
\plotone{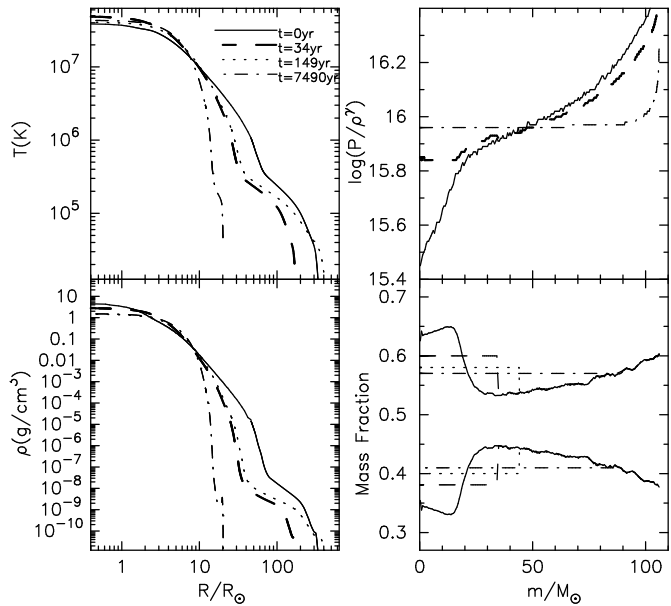}
\caption{Evolution of 
stellar structure of UE1. While corresponding to 
Figure \ref{fig:str1_rhot} for EQ1, this figure focuses on the initial 
thermal adjustment phase. The solid, dashed, dotted, and dot-dashed lines 
are the results at $t=0$, 34 yr, 149 yr, and 7490 yr, respectively, where 
in the top-right panel the dotted line is omitted to avoid confusion. 
Followed by the initial contraction before $t<34$ yr, the star slightly 
expands between 34 yr $< t < 149$ yr and again contracts between
149yr $< t < 7490$ yr to the minimum radius.}
\label{fig:str2_rhot}
\end{figure}

\begin{figure}
\epsscale{0.9}
\plotone{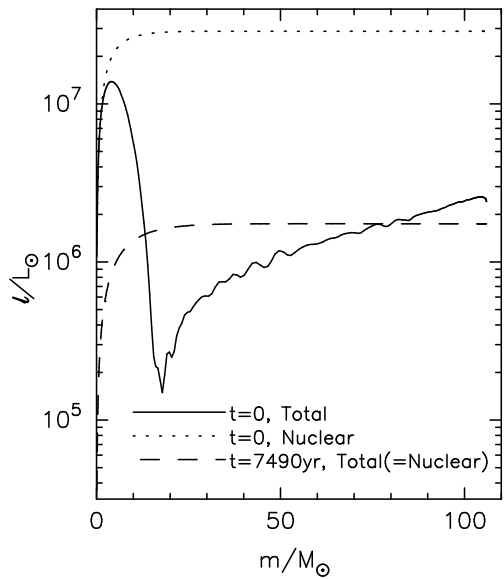}
\caption{
Luminosity normalized by $L_{\odot}$ on $m/M_{\odot}$ of UE1. 
The dotted line is the luminosity due to the nuclear burning 
at $t=0$. The solid and dashed lines are the total luminosity at $t=0$ 
and 7490 yr, respectively, whereas the luminosity at $t=7490$ yr is 
from the nuclear energy. 
}
\label{fig:str2_Lr}
\end{figure}


The evolution of the merger products of the unequal mass stars is 
different from that of the equal mass cases during the initial 
contraction phase, while the later evolution follows chemically 
homogeneous single stars with the corresponding masses and abundances in 
both cases. 
Figures \ref{fig:str2_rhot} and \ref{fig:str2_Lr} present the results of 
UE1, which correspond to Figures \ref{fig:str1_rhot} and \ref{fig:str1_Lr} 
for EQ1. 
As we have shown in \S \ref{sec:sph}, the most important difference 
is that the interior is not well-mixed (Figures \ref{fig:SPH_SNAP_UE} 
and \ref{fig:str2_ue}). Just after the collision, 
the lower mass parent star (Star 2) sinks to the center without 
sufficient mixing because it has a higher density and lower entropy.  
This star is more H-rich since the nuclear burning proceeds more slowly than 
in the massive partner (the bottom panel of Figure \ref{fig:str2_ue}). 
Therefore, the merged star consists of a H-rich core and a 
He-rich outer region as shown in the bottom right panel of 
Figure \ref{fig:str2_rhot}. 
Reflecting the higher density in Star 2, the density of the core
of the collision product becomes slightly larger than the thermal 
equilibrium value, while the lower density envelope extends to the outer 
region (the bottom-left panel of Figure \ref{fig:str2_rhot}).  

Due to the high density as well as the moderate temperature, the nuclear 
burning takes place rather rapidly even just after the 
merging event in the unequal mass case 
(Figure \ref{fig:str2_Lr}).  
The energy release rate by the nuclear reaction exceeds 
the total luminosity because the nuclear energy is also used to increase 
the temperature (internal energy; the top-left panel of Figure 
\ref{fig:str2_rhot}) and to expand the 
core (work on gas, i.e., the decrease of the core density; see the 
bottom-left panel of Figure \ref{fig:str2_rhot}). 
Accordingly, the envelope also expands from $t=34$ yr to $149$ yr 
(Figures \ref{fig:t-r1} and \ref{fig:str2_rhot}). 
In fact, at $t=149$ yr, the radius becomes, $R_{\star} = 412R_{\odot}$, 
which is larger than $R_{\star} = 337R_{\odot}$ just after the merger ($t=0$). 
Reflecting the initial expansion, the evolutionary path in the HR diagram  
(Figure \ref{fig:hr1}) is also more complicated compared to the 
equal mass case. 

During this phase, the size of the convective core is small, $m < 
35M_{\odot}$, because the entropy 
(the top-right panel of Figure \ref{fig:str2_rhot}) stays low in the core.
Therefore, the H-rich core is still preserved without mixing 
with the outer region (the bottom right panel of Figure \ref{fig:str2_rhot}).  
Incidently, the entire region outside $m 
= 17M_{\odot}$ becomes convectively stable with 
respect to the Schwartzschild criterion, $d(p/\rho^\gamma)/dr > 0$ 
(the top-right panel of \ref{fig:str2_rhot}). 
Between $17M_{\odot}< m < 35M_{\odot}$, the gradient of mean 
molecular weight, $\mu$, leads to mixing because heavier He is more abundant 
in the upper layer; this region is unstable only by the Ledoux 
criterion. Note that this is opposed to usual situations, in 
which heavier elements are more abundant in a lower region and $\mu$ 
gradient contributes to stabilization.

The initial expansion between $34<t<149$ yr is followed by the usual 
contraction to the equilibrium state through thermal adjustment.
The chemical abundance becomes homogeneous from inward as the convective 
core grows to $m\simeq 80M_{\odot}$.  
The minimum radius, $R_{\star} = 20.1 R_{\odot}$, occurs at $t=7490$ yr, 
roughly corresponding to  
$\tau_{\rm KH}$. The later evolution traces the evolution of the single 
homogeneous star, which is the same as in the equal mass case. 
The duration of the MS ($\approx 1.6$Myr) is again shorter than 
the corresponding lifetime ($\approx 2.5$Myr) for a solar abundance star. 

The evolution of UE2 is similar to the evolution of UE1 (Figure \ref{fig:hr1}): 
The merger product initially consists of a H-rich core and a He-rich 
envelope. Although this structure is maintained at first due to the small 
convective core, the interior becomes homogeneous after $t\gtrsim 5000$ yr 
as the convective core grows. The later evolution resembles the evolution 
of the corresponding single homogeneous star.  

\subsubsection{Mass Loss by Stellar Winds}
\label{sec:ml}

\begin{figure}
\epsscale{0.85}
\plotone{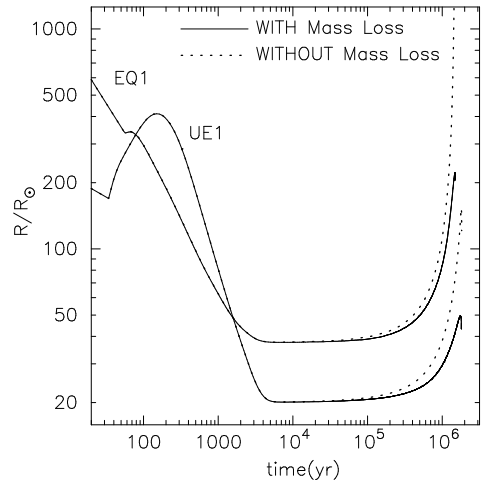}
\caption{Evolution of the stellar radii of EQ1 and UE1 that take into account 
mass in the stellar evolution (solid lines), in comparison with the results 
without mass loss (dotted lines). 
}
\label{fig:t-r2}
\end{figure}

\begin{figure}
\epsscale{0.85}
\plotone{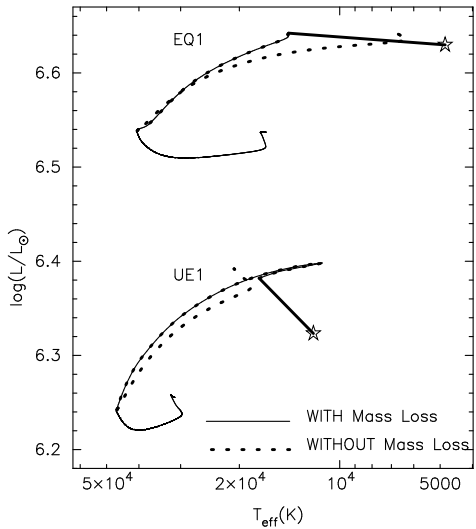}
\caption{Evolution of EQ1 and UE1 with (solid lines) and without (dotted lines) 
mass loss in a HR diagram. 
}
\label{fig:hr2}
\end{figure}

So far we have not considered the effect of mass loss by radiation-driven 
stellar winds \citep{cac75}. However, it is supposed to affect the evolution 
of the merged massive stars. 
We study the evolution of EQ1 and UE1 by explicitly taking into account mass 
loss in the stellar evolution calculations.
Here we have adopted
the mass loss rate $\dot{M}$ of solar metallicity gas from 
\citet{kud02}, which tabulates $\dot{M}$ as a function of $L_{\star}$,  
$T_{\rm eff}$, and metallicity\footnote{Although in \citet{kud02} the 
dependence of $\dot{M}$ on He abundance, $Y$, is not explicitly presented, 
observation of Wolf-Rayet stars shows that $\dot{M}$ has a  
dependence on $\propto Y^{1.73}$ \citep{nug00}. Our merger products 
are He-rich $Y\approx 0.4$ in the envelopes, compared to the Sun ($Y=0.28$), 
hence, $\dot{M}$ could be be larger by a factor of 1.5-2 than that by 
\citet{kud02}. However, even if $\dot{M}$ becomes larger by this extent, 
we suppose that the effect of the mass loss is not still crucial during the 
MS phase. }. 

In Figures \ref{fig:t-r2} \& \ref{fig:hr2}  we show the evolution of EQ1 and 
UE1 with mass loss, in comparison with the results without mass loss.
Our collision products have typically $\dot{M} \sim (1-3)\times 
10^{-5}$ $M_{\odot}$/yr, and, $\sim 10 - 30M_{\odot}$ is lost during the 
MS phase of $\sim$1-2Myr. 
The differences of the stellar radii between the cases with mass loss and 
the cases without mass loss are not large (Figure 
\ref{fig:t-r2}); they are less than 10\% except at the very end of the MS
phase ($t\gtrsim 1$Myr in EQ1 and $t\gtrsim 1.5$Myr in UE1).
Figure \ref{fig:hr2} illustrates that the luminosity becomes slightly smaller 
by $\approx 0.1$dex at later epochs because of the mass loss.

Once runaway collisions start in a dense cluster, the timescale of subsequent 
collisions is much shorter than 1Myr \citep{pzetal04}.   
Therefore, further stellar collisions of the merged stars 
would take place before the mass loss becomes important 
in the stellar evolution. 
Metal-poor stars give even smaller mass loss rates than solar abundance 
stars.
Thus, we can conclude that stellar mass loss does not stop the increase in mass due
to runaway collisions, provided that the 
metallicity is comparable to or smaller than the solar value.




\section{Conclusion and Prospect}
Bearing in mind formation of supermassive stars and IMBHs in dense star 
clusters, we have studied the hydrodynamical processes during collisions 
of massive stars and the evolution of the merger products.  
After the collisions, the merged stars settle down to dynamically stable 
states on typical timescales of days to weeks, well before they would undergo
further collisions.
During the merger events, the stars typically lose $\sim 10$\% of the total 
mass. 
The merger products are He-rich because of the nuclear burning of their 
parent stars. 
The interior of the merger product of equal-mass progenitors is 
well-mixed during this dynamical phase because of the low entropy barrier.
On the other hand, during the merging of unequal-mass stars, 
the less massive star sinks into the core, and the more massive
partner is elongated by tidal interaction to finally form the envelope. 
Since the nuclear burning 
took place slower in the less massive progenitor, the merged star consists of 
an H-rich core and an He-rich envelope.   

After the merger phase, the merged stars evolve to thermal 
equilibrium states on Kelvin-Helmholtz timescales, $10^{3-4}$ yr. 
The evolution of the collision product of equal mass stars is very 
similar to a pre-main sequence star; the star monotonically contracts and the 
luminosity is mainly supplied from the release of gravitational energy. 
On the other hand, the evolution of the merger product of unequal mass 
stars is rather complicated due to the poorly mixed interior; 
the nuclear burning is already switched on owing to the sufficiently dense 
core, and as a result, the star slightly expands at first, which is followed 
by the usual contraction. 

Just after the merging the radius is larger than
the equilibrium value by a factor of 10-100, while it goes down
to a few times the equilibrium value in less than 1000 years (See
figure \ref{fig:t-r1}). N-body simulations that assume the mass-radius 
relation for MS stars show that collisions typically take place every $\sim 
3\times 10^4$yr in a very dense region (Baumgardt et al. 2006). 
The collision probability will be enhanced 
when we take into account such realistic stellar radii. 
Then, a small fraction of merged stars might experience further 
collisions during the initial contraction phase if they are in a very 
dense region. 
However, most of the collisions are off-axis, and 
we expect that such off-axis collisions simply blow away a tiny 
fraction of the outer envelope, rather than resulting in a merger, 
because its density is quite low (Figures \ref{fig:str1_rhot} and 
\ref{fig:str2_rhot}).


After the thermal adjustment phase, the merged stars enter a stable nuclear 
burning phase and their evolution is well approximated by those of single 
homogeneous stars with corresponding masses and abundances. 
An important point here is that the lifetimes of merger products are 
shorter than solar abundance stars with the same masses because they
are already He-rich from the beginning.    
\citet{pzetal04} and \citet{bau06} assumed 3 Myr as the lifetime of runaway 
stars for the stellar dynamics simulations.  
Our result have shown that massive stars typically collide after 1Myr from 
cluster formation and that the lifetime of the merged stars is $\sim$2Myr; 
these results confirm that the assumption of 3Myr is quite reasonable. 
This is robust even if merged stars experience further collisions because 
the collision products become more He (or heavier element)-rich and their 
lifetime is short.     

Our simulations show that neither mass loss during stellar collisions 
nor mass loss by the stellar winds prevents the growth in mass of the 
collision products. 
We can therefore anticipate that the scenario of the formation of 
supermassive stars by successive collisions \citep{ebi01,pzetal04} is 
likely to occur in realistic situations. 
Because of the nuclear burning, the merged stars become more H-poor.  
Finally, the lifetimes of massive He-rich descendants are much shorter 
than those of solar abundance stars with corresponding masses. 
We speculate that the very end-products of runaway 
collisions would form IMBHs quickly before suffering substantial mass loss; 
the key is that the material which is finally taken in a 
supermassive star spends most of the time in less massive stars, which are 
not influenced by mass loss so much. Therefore, supermassive stars are 
possibly formed by successive collisions, although it seems difficult through 
the evolution of very massive single stars \citep{bel07}.  
However, our present work does not quantitatively treat this final process.   
For such purpose we need to the study evolution of very massive 
($\sim 1000M_{\odot}$) and chemically evolved (He-rich with 
abundance gradient) stars. 
We plan to carry out such simulations in the future.

The gravity calculation of the SPH simulations has been done
with reconfigurable computing board PROGRAPE-3.
NN would like to thank Dr. T.Hamada for discussions and help
regarding gravity calculations on PROGRAPE-3.
This work is supported in part by a Grant-in-Aid for Scientific
Research (19015004 : TKS) from the Ministry of
Education, Culture, Sports, Science, and Technology of Japan.
HB acknowledges support from the Japan Society for the Promotion of Science
through Short-term visitor grant S-06709.

\begin{appendix}

\section{Dependence on Numerical Resolution}
\label{Ndep}
In the present work, we choose rather small numbers of particles 
($N \sim 10,000 - 20,000$) in Step (2).
In previous extensive simulations of colliding stars by \cite{Freitag_2005}, 
they have repeated the same runs with different N's to see how the numerical 
resolution affects the hydrodynamical simulations of colliding stars.
They have concluded that $N \sim 10,000 - 30,000$ per star is adequate 
to construct comprehensive tables for simulations of dense stellar systems. 
Our main interest in the present work, which is different from theirs,
is the fate of collision products of massive stars as a result of the 
stellar evolution. 

\begin{figure}
\epsscale{1.0}
\plottwo{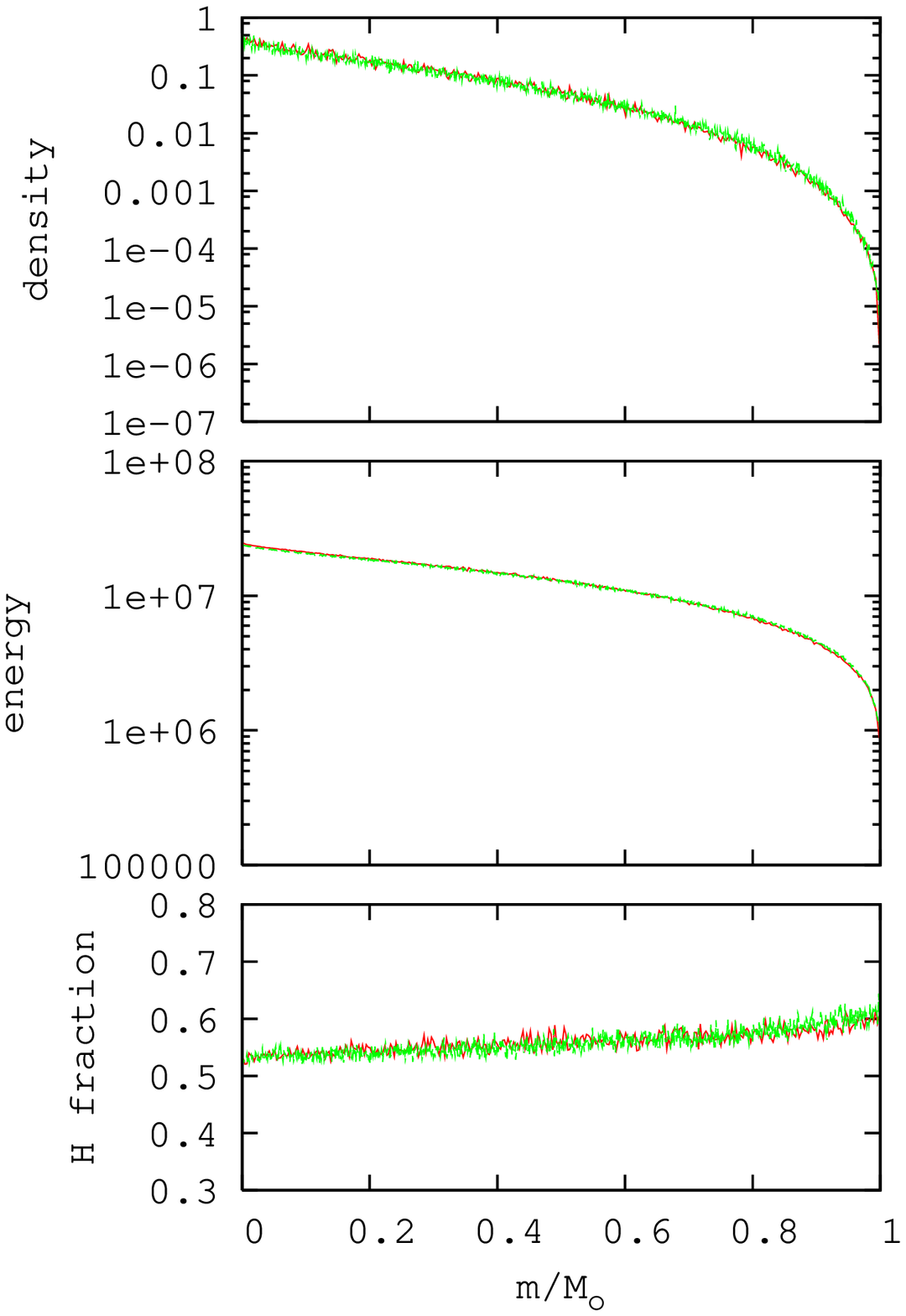}{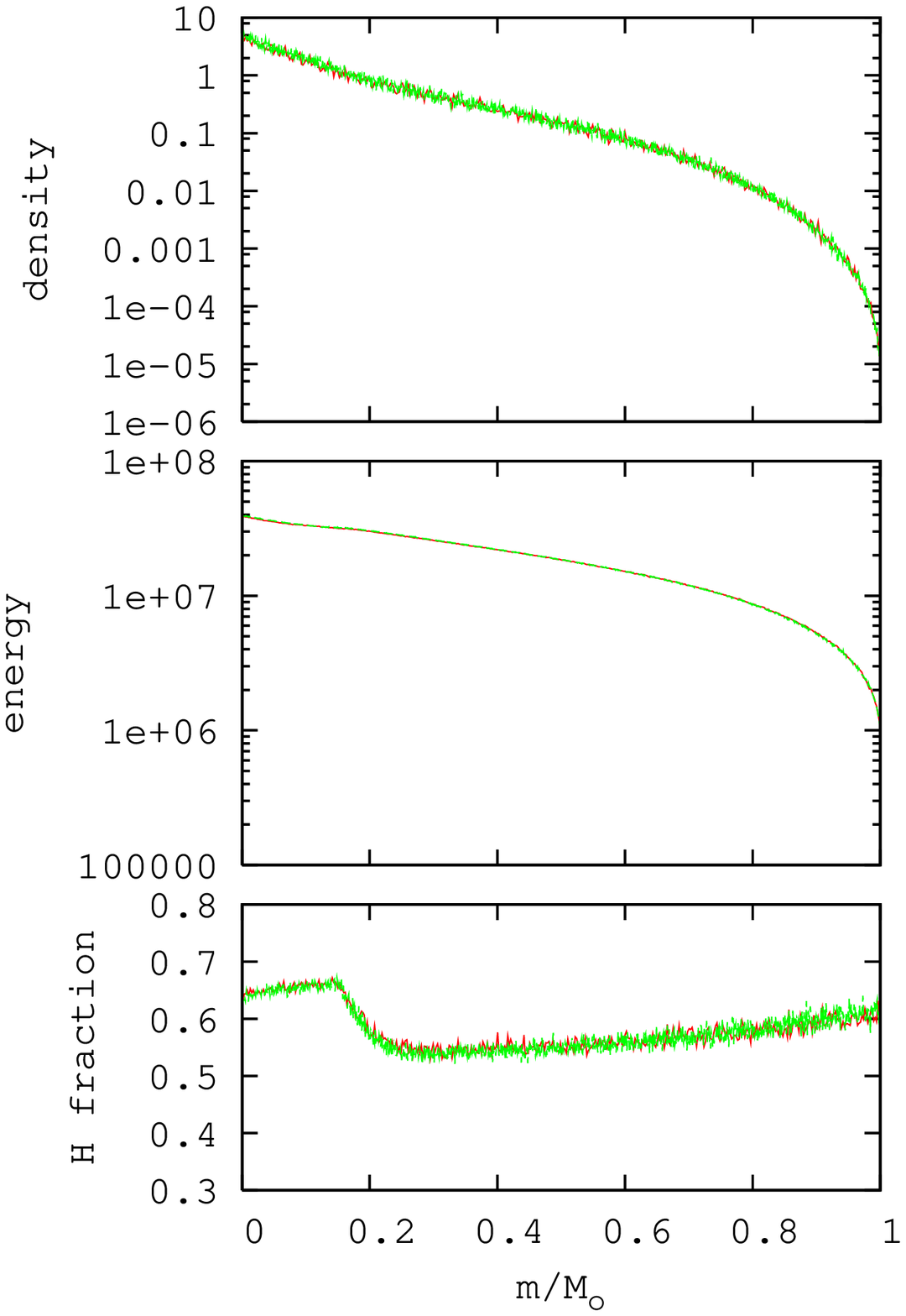}
\caption{
Structures of the merged stars EQ1 (left) and UE1 (right) obtained with SPH 
simulations using different $N$.
From top to bottom, density, internal energy and H abundance are plotted 
as functions of mass radius, $m/M_{\odot}$.
Red and green lines correspond to results for $N = 20,000$ and $60,000$ in 
EQ1 and $N = 26,305$ and $78,915$ in UE1, respectively.
}
\label{fig:eq1comp}
\end{figure}

The results of the resolution tests for EQ1 and UE1 runs
are shown in Figure \ref{fig:eq1comp}. 
All the quantities are averaged over every 55 particles in mass coordinate.
In both cases, the runs with three times more N shows almost identical results
to the corresponding original runs.
Moreover, we practically average the quantities of SPH simulations of Step (2) 
to give 1D spherical symmetric structure used in Step (3). 
Then, fine details in the results are smoothed out and the tiny 
differences seen in Figure \ref{fig:eq1comp} 
have little impact on calculations in Step (3).
We conclude that the relatively small N that we are using in the present work 
is sufficient for our purpose.

\end{appendix}

\end{document}